\documentclass[aps,prl,twocolumn,superscriptaddress,showpacs,floatfix,longbibliography]{revtex4-2}
\usepackage{amsmath,amssymb,amsfonts,float,graphics,epsfig,epstopdf,color,verbatim,tabularx,bm,multirow,appendix,hyperref}
\usepackage{lmodern}
\usepackage{ytableau}
\usepackage{color}
\usepackage{bm}

\newcommand{\beq} {\begin{equation}}
	\newcommand{\eeq} {\end{equation}}
\newcommand{\bea} {\begin{eqnarray}}
	\newcommand{\eea} {\end{eqnarray}}
\newcommand{\be} {\begin{equation}}
	\newcommand{\ee} {\end{equation}}

\newcommand{\mysection}[1]{\noindent \textit{\textbf{#1}}}

\begin{document}
	
	\title {Polariton-Induced Unconventional Superconductivity and Emergent SU(2) Symmetry in Moir\'e flat bands}

	\author{Kai Sun}
	\email{sunkai@umich.edu}
	\affiliation{Department of Physics, University of Michigan, Ann Arbor, Michigan 48109, USA}

	\author{Hui Deng}
	\email{dengh@umich.edu}
	\affiliation{Department of Physics, University of Michigan, Ann Arbor, Michigan 48109, USA}

	\date{\today}
	
	\begin{abstract}

We propose a polariton-moir\'e coupled system to realize an exact solvable model of strongly correlated superconductors. 
The polariton condensate induces intervalley attraction between moir\'e electrons in quasi-flat topological bands, leading to emergent SU(2) symmetry and exactly solutions of the many-body ground state. 
This system expands the condensate-induced superconductivity to the strong-correlation regime, enabling superconductivity at elevated temperatures and allowing the study of non-Fermi liquid states with a solvable, predictive model. It is promising as a versatile platform for emulating unconventional superconductivity and other strongly correlated phenomena in complex, correlated materials. 

	\end{abstract}
	
	\maketitle
	
Microcavity exciton-polaritons~\cite{weisbuch_observation_1992}, which are hybridized modes of excitons and cavity photons, have long been recognized as a conduit for bosonic many-body quantum states in solids~\cite{littlewood_optical_2007,deng_exciton-polariton_2010,carusotto_quantum_2013,byrnes_excitonpolariton_2014,amo_exciton-polaritons_2016,kavokin_polariton_2022}. Thanks to the experimental accessibility through their photonic constituent, polaritons were the first system where the second-order correlation function of a Bose-Einstein condensate (BEC) was measured~\cite{deng_condensation_2002}, and the first \textit{solid-state} system to host a BEC~\cite{deng_condensation_2002,kasprzak_boseeinstein_2006,balili_bose-einstein_2007}. Intriguingly, such polariton condensates have been predicted to induce superconductivity in an electron gas, where quantum fluctuations around the condensate lead to attractive interactions and pairing between electrons~\cite{Laussy2010,Shelykh2010,Cotlet2016,Villegas2019,Julku2022}. 
Furthermore, it has been proposed that valley degrees of freedom in 2D materials could even stabilize topological Fulde–Ferrell–Larkin–Ovchinnikov (FFLO) states~\cite{Julku2022}.

The polariton-induced superconductors studied in these pioneering works have been confined to the weak correlation limit. In this limit, the Bardeen Cooper Schreiffer (BCS) and Eliashberg theories are applicable and provide accurate predictions of $T_c$, up to a few Kelvin for real-world parameters~\cite{Cotlet2016}. Reaching a high $T_c$ may require going beyond the weak-correlation limit. 
On the other hand, the study of strongly correlated phenomena, such as pseudo-gap and unconventional superconductivity in high $T_c$ cuprates, is fraught with unresolved fundamental questions, due to the complexity of the materials (e.g., chemical doping) along with their strong correlation nature. However, recent theoretical studies have revealed a well-controlled pathway to realize and study these phenomena via exactly solvable models, such as attractive Hubbard (or similar) models in flat or topological flat bands~\cite{PhysRevB.94.245149, PhysRevB.102.201112, Xu2022, herzogarbeitman2022manybody}. Such models allow for exact solutions of the many-body systems and thus precise predictions of quantum phases and physical observables.
Yet realization of such models requires flat bands and attractive on-site Hubbard interaction, both of which are very difficult in conventional electronic systems. 

In this letter, we propose a system to implement an exactly solvable model for non-Fermi liquids using a polariton condensate coupled to a quasi-flat-band moir\'e electron gas. 
We make use of a recent study that shows 
the on-site attractions in previous models can be generalized to systems with intervalley attractions of any functional form~\cite{Xu2022}. Polariton condensate in the proximity of the electron gas can induce such attractions~\cite{Laussy2010,Shelykh2010,Cotlet2016,Villegas2019,Julku2022}. At the same time, topological flat bands have been theoretically predicted~\cite{Devakul2021,HeqiuLi2021,PhysRevB.107.L201109, Wang2023,reddy2023fractional} and experimentally demonstrated~\cite{Cai2023, Zeng2023, park2023observation} in twisted homobilayers of transition metal dichalcogenides (TMDs).
We show that the proposed polariton-moir\'e system expands the condensate-induced superconductivity to the strong-correlation regime. Superconductivity at elevated temperatures and non-Fermi liquid states above the superconducting dome are predicted by an exact sovolable model. 
Notably, both polariton and moir\'e materials are relatively simple to model and easy to tune. 
Their combined system avoids many of the complexities of correlated materials and provides a versatile and controllable platform for studying strongly correlated phenomena.
\begin{figure}
		\includegraphics[width=0.4\textwidth]{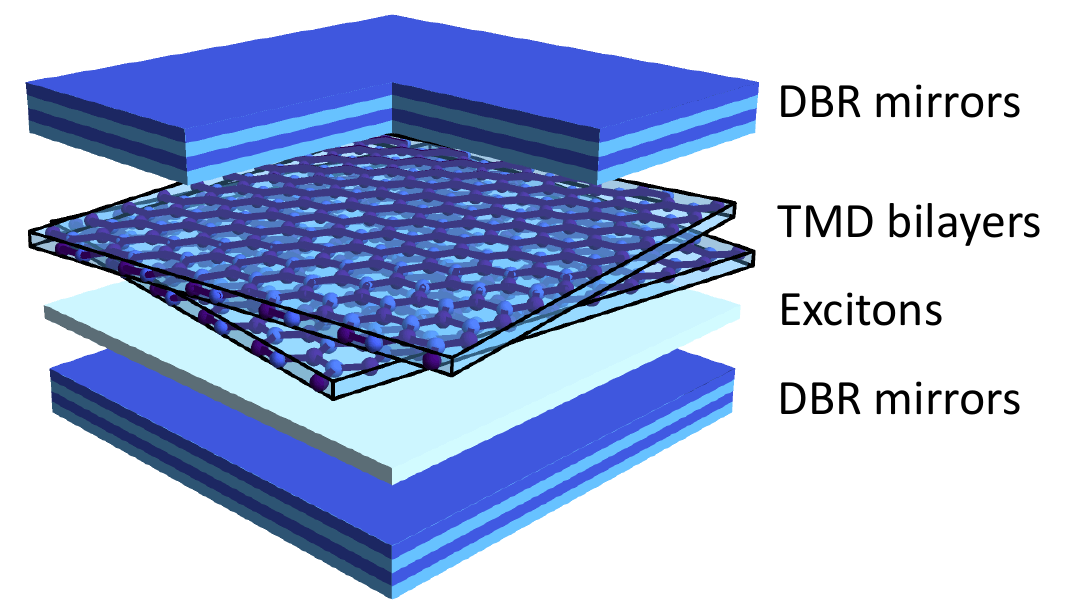}
		\caption{
Illustration of a coupled polariton-moir\'e system that realizes an exactly-solvable model of strongly-correlated superconductors. A pair of twisted TMD monolayers form a moir\'e lattice with nearly flat moir\'e bands. The moir\'e electrons interaction with excitons in an insulating TMD monolayer nearby. A cavity encloses both and couples strongly with the excitons to form polaritons. An external laser can create a polariton condensate that induces attractive interactions between the moir\'e electrons. 
  }
\label{fig:setup}
\vspace{-12pt}
	\end{figure}
\begin{figure}
		\includegraphics[width=0.4\textwidth]{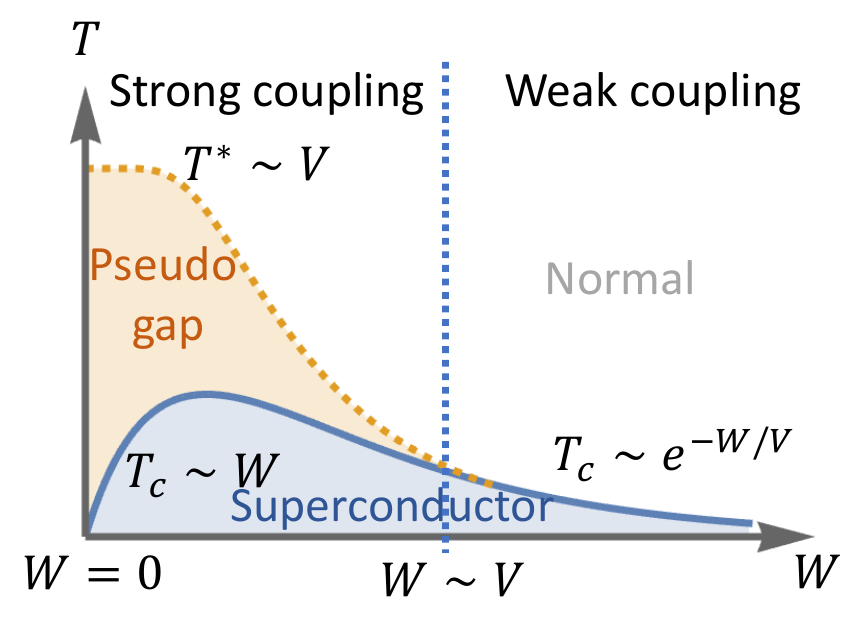}
		\caption{Phase diagram of the coupled polariton-moir\'e system, showing the pseudogap temperature ($T^*$ dashed line) and the superconducting transition temperature ($T_C$ solid line), as functions of the flat-band bandwidth $W$ compared to the intervalley attraction $V=\tilde{V}^{eff}$. By adjusting the twist angle, $W$ can be precisely controlled, while $\tilde{V}^{eff}$ is induced and controlled by a polariton condensate. In the weak-coupling regime ($\tilde{V}^{eff} \ll W$), BCS theory applies. In the strong coupling realm ($\tilde{V}^{eff} \gg W$), an exact solution emerges at $W=0$. Near this limit, $T_C \sim W$ and $T^* \sim \tilde{V}^{eff}$. Between $T^*$ and $T_C$, a pseudogap phase emerges, where Cooper pairs are formed but fail to condense -- a key signature of strongly-correlated superconductors.
  }
  \vspace{-12pt}
\label{fig:bandwidthvsTc}
	\end{figure}
 
\mysection{Physical system.}
As illustrated in Fig.~\ref{fig:setup}, our proposed setup consists of an electrically gated, twisted TMD homobilayer and a nearby insulating 2D material with a smaller bandgap, both enclosed in an optical cavity. The homobilayer has a twist angle near the magic angle for the topological flat bands; gating controls doping and partially fills the flat bands. The insulating 2D material layer, such as a monolayer TMD, is positioned at the antinode of the cavity and strongly couples with the cavity to form polaritons; it hosts a polariton condensate, either spontaneously formed or injected. Quantum fluctuations around the condensate induce intervalley attraction among electrons in the twisted bilayer~\cite{Laussy2010,Shelykh2010,Cotlet2016,Villegas2019,Julku2022}. 

\mysection{Model.}
To model the above system, we consider zero doping in the insulating 2D material and neglect excitonic effects in the moir\'e structure. Thereafter, electrons refer to electrons in the moir\'e lattice, and excitons refer to excitons in the insulating layer. Then the system is described by the following Hamiltonian:
\begin{align}
H=H_{0}^{e}+H_{0}^{p}+H_{I}^{e-ex}+H_{I}^{ex-ex}+H_{I}^{e-e}.
\end{align}
Here $H_{0}^{e}$ and $H_{0}^{p}$ correspond to the non-interacting Hamiltonian for the electrons and polaritons, respectively, 
$H_{I}^{e-ex}$ is the electron-exciton coupling, and $H_{I}^{ex-ex}$ is the exciton-exciton interaction. Detailed forms of these first four terms are provided in Appendix A. 
The last term, $H_{I}^{e-e}$ describes electron-electron interactions. Its dominant part is the repulsive density-density interaction:
\begin{align}
\label{eq:H_I_ee}
H_{I}^{e-e}=\frac{1}{2}\sum_{q} V^{Coul}_q :\rho_{q}\rho_{-q}:
\end{align}
where $:\ldots:$ signifies normal ordering and $\rho_{q}=\rho_{\tau,q}+\rho_{-\tau,q}$ aggregates the total fermion density from both valleys, with $\tau =\pm K$ denoting the valley index.
The interaction strength $V^{Coul}_q$ can be tailored to the experimental conditions, considering Coulomb, screened-Coulomb, or gate-screened Coulomb interactions.

As will be discussed below, this model system realizes a correlated superconductor that can be exactly solved in the strong-correlation limit. The schematic phase diagram of this system is depicted in Fig.~\ref{fig:bandwidthvsTc}. 

\mysection{Mediated interaction and effective model.}
Previous work~\cite{Laussy2010,Shelykh2010,Cotlet2016,Villegas2019,Julku2022} has shown that, through electron-exciton and exciton-exciton interactions ($H_{I}^{e-ex}$ and $H_{I}^{ex-ex}$), a polariton condensate, encapsulated in $H_{0}^{p}$, induces attractions between both intravalley and intervalley electrons. Such attactions may lead to two distinct superconducting states with intervalley~\cite{Cotlet2016} and intravalley~\cite{Julku2022} pairings, respectively.

	\begin{figure}
		\includegraphics[width=0.4\textwidth]{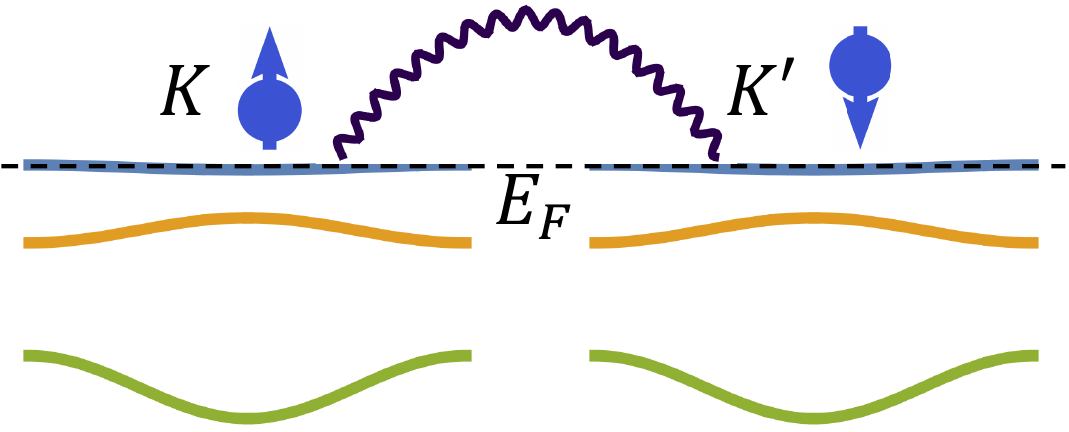}
		\caption{Illustration of moiré minibands and polariton-induced attractions: The solid lines correspond to the minibands in the K and K' valleys, carrying opposite spins. Near the magic twisting angle, the top miniband is near-flat. The dashed line indicates the Fermi energy, while the oscillating line illustrates intervalley attractions mediated by a polariton condensate.}
\label{fig:minibands}
\vspace{-12pt}
	\end{figure}

For the setup depicted in Fig.~\ref{fig:setup}, the intervalley attraction dominates, while intravalley interaction manifests self-cancellation between direct (Hartree) and exchange (Fock) contributions due to the Pauli exclusion principle. 
Specifically, the intravalley interaction between the electrons is:
\begin{align}
H_{intra}=\frac{1}{2}\sum_{k,k',q} V_q \psi^\dagger_{k+q,\tau}\psi^\dagger_{k'-q,\tau}\psi_{k',\tau}\psi_{k,\tau}.
\label{eq:intra:interaction}
\end{align}
Due to strong spin-orbit coupling in TMDs, electron spins are fully polarized. Hence there are no internal degrees of freedom for fermions in each valley. By swapping the two fermion annihilation operators in Eq.~\eqref{eq:intra:interaction}, we obtain a negative-coefficient interaction term, the exchange term, stemming from anti-commutation. This negative term partially cancels the original (direct) term. At $k-k'=0$, Pauli exclusion principle requires $\psi_{k',\tau}\psi_{k,\tau}=0$, or complete cancellation of the intravalley interactions. For small but non-zero $k-k'$, cancellation is imperfect, with a residual interaction strength $V_{intra}=(V_q-V_{q+k-k'})/2 \propto |k-k'|$, as detailed in Appendix B.

For a Fermi liquid, the main contributions to interactions are from fermions in proximity to the Fermi wavevector $k_F$. This is a fundamental characteristic of Fermi liquids, supported by the renormalization group analysis~\cite{Shankar1994}. Consequently, the most significant contributions to Eq.~\eqref{eq:intra:interaction} arise from terms satisfying $|k-k'|\lesssim 2 k_F$. At low doping, characterized by a small electron density $\rho_e$, it follows that $V_{intra}\propto k_F \propto \sqrt{\rho_e}$ and $V_{intra}$ vanishes as $\rho_e$ approaches zero. In contrast, interval interactions evade such substantial cancellation and we have $V_{inter} \gg V_{intra}$ at low doping. 

Considering only the intervally terms, the effective interactions between electrons becomes:
\begin{align}
H_{int}=-\frac{1}{2}\sum_{q,\tau} |V^{eff}_q| \rho_{q,\tau} \rho_{-q,-\tau},
\label{eq:int:intervalley}
\end{align}
with
\begin{align}
V^{eff}_{q}=V^{Coul}_q-\frac{2 \rho_0(g_{e-ex} C_0 C_k)^2 (\epsilon_q^{LP}-\epsilon_0^{LP})}{\omega_n^2+E_q^2} <0
\end{align}
The first term, $V^{Coul}_{q}$, is the repulsive screened or gate-screened Coulomb interaction in Eq.~\ref{eq:H_I_ee}. The second term is the polariton-induced attraction~\cite{Camacho-Guardian2021, Julku2022}, where $\rho_0$ is the polariton condensate density, $\epsilon_q^{LP}$ and $E_q$ are the energies of the lower polariton and Bogoliubov excitations, $\omega_n$ is the Matsubara frequency, $g_{e-ex}$ is the electron-exciton interaction coefficient, and $C_k$ is the Hopfield coefficients~\cite{Hopfield1958}. To higlight that $V_q^{eff}$ is an attraction, 
we explicitly show the negative sign in Eq.~\eqref{eq:int:intervalley} by taking the absolute value of $V_q^{eff}$.

To align with the solvable model detailed in Ref.~\cite{Xu2022}, we rewrite $H_{int}$ in a total square form:
\begin{align}
H_{int}=\sum_{q} \frac{|V^{eff}_{q}|}{2}  (\rho_{q,\tau}-\rho_{q,-\tau})(\rho_{-q,\tau}-\rho_{-q,-\tau}).
\label{eq:int:before:projection}
\end{align}
The added intravalley repulsion, $\rho_{q,\tau}\rho_{q,\tau}$,
is inconsequential due to the cancellation discussed above. Hence, Eq.~\eqref{eq:int:before:projection} preserves the essential physics of the system. 

Importantly, the interaction term expressed in Eq.~\eqref{eq:int:before:projection} satisfies the requirement to realize the solvable model in Ref.~\cite{Xu2022}. Even beyond the exact flat-band limit, this model circumvents the sign problem in quantum Monte Carlo simulations, allowing the determination of its phase diagram~\cite{Xu2022}.

\mysection{SU(2) symmetry and exact solutions.}
Next we show emergent SU(2) symmetry in the flat-band limit of our system, leading to exact solutions of the system. As illustrated in Fig.~\ref{fig:minibands}, we consider twisted TMD bilayers near a magic angle, where two degenerate flat bands emerge (one for each valley), carrying opposite spins. We focus on the scenario when the flat bands are partially filled, with Fermi energy aligned to the flat bands. In this setup, other moir\'e bands can be neglected, as they are far from the Fermi energy. Consequently, the electron Hamiltonian projected to the flat band adopts the following form:
\begin{align}
H=\sum_{\tau,k}\epsilon_k c_k^\dagger c_k +H_{int},
\label{eq:full:Ham}
\end{align}
where $c^\dagger_k$ denotes the fermion operator generating a Bloch wave of the moir\'e flat band, while $\epsilon_k$ represents the dispersion of this band. 
Once projected, the density operator in $H_{int}$ becomes
\begin{align}
\rho_{q,\tau}= \sum_k \lambda_{\tau,q,k}c^\dagger_{k,\tau} c_{k+q,\tau}
\label{eq:density:projected}
\end{align}
where $\lambda_{\tau,q,k}=\langle u_{\tau,k} | u_{\tau,k+q} \rangle$ and $|u_{\tau,k} \rangle$ is the Bloch eigenstates of the flat band determined by $H_0^e$~\cite{Lee2019, HeqiuLi2021}.

	\begin{figure}
		\includegraphics[width=0.5\textwidth]{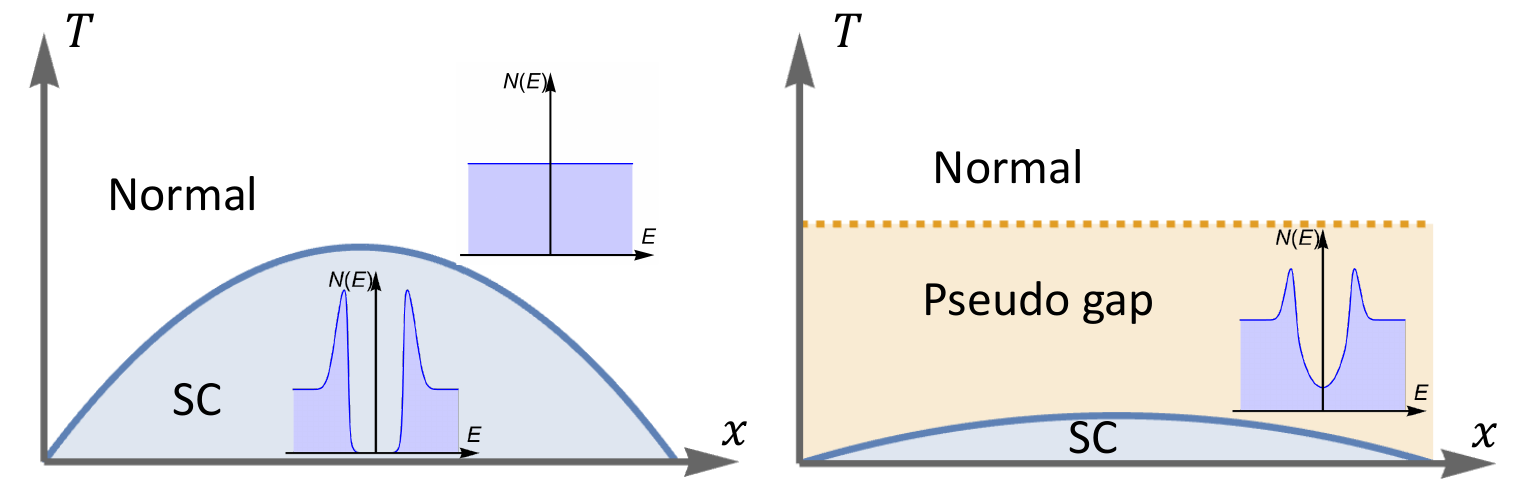}
		\caption{Comparison of the temperature-density phase diagrams in the weakly (left panel) and strongly correlated (right panel) regimes. 
  The density $x$ can be represented by the fermion filling factor ($0<x<1$). Note that the vertical axis (temperature) scales differently in the two panels, so direct $T_c$ comparison is unsuitable. Insets show the fermion density of states near Fermi energy.
  Left, weakly-correlated BCS-type superconductors. 
  Right, strongly-correlated superconductors, where the dashed line depicts the pseudogap temperature ($T^*$), while the $T_c$ for the superconducting dome scales with the band width $T_c\sim W$ and vanishes in the exact flat band limit of $W=0$.}
\label{fig:phasediagram}
\vspace{-9pt}
	\end{figure}

In the exact flat-band limit, characterized by a constant dispersion $\epsilon_k$, the first term in Eq.~\eqref{eq:full:Ham} contributes a trivial constant and can be ignored. Consequently, the Hamiltonian simplifies to $H=H_{int}$. A pivotal characteristic emerges in this Hamiltonian: an emergent $SU(2)$ symmetry~\cite{Xu2022}. This symmetry remains in effect regardless of $V^{eff}_{q}$ being short- or long-range.

We define this SU(2) symmetry by introducing three operators:
\begin{align}
\Delta^\dagger&=\sum_k c^{\dagger}_{k,\tau}c^{\dagger}_{-k,-\tau}=S^{+}\\
\Delta&=\sum_k c_{-k,-\tau}c_{k,\tau}=S^{-}\\
N&=\sum_{k,\tau}  (c^{\dagger}_{k,\tau}c_{k,-\tau}-1/2)=S_z.
\end{align}
Here, $\Delta^\dagger$ ($\Delta$) creates (annihilates) a Cooper pair, with its expectation value constituting the order parameter for intervalley zero-momentum pairing. $N$ represents the fermion density operator, referenced from the half-filling. Importantly, these three operators follow the same commutation relations as the spin raising/lowering operators $S^{\pm}$ and $S_z$, thus embodying the su(2) algebra structure. More significantly, these three operators commute with $H_{int}$, thus establishing a SU(2) symmetry group of the system in the flat band limit.

This SU(2) symmetry unveils a compelling insight: introducing a Cooper pair incurs no energy cost in this system. Let $|\psi\rangle$ denote a many-body eigenstate of the many-body Hamiltonian $H |\psi\rangle = E |\psi\rangle$. The commutation relation $[H,\Delta^\dagger]=0$ implies that $H\Delta^\dagger |\psi\rangle = E \Delta^\dagger |\psi\rangle$, rendering $\Delta^\dagger|\psi\rangle$ a degenerate eigenstate of $H$ with an identical eigenenergy. Therefore, since the $\Delta^\dagger$ operator introduces a Cooper pair, adding a Cooper pair to the system has no energy cost. This observation extends to adding any number of Cooper pairs, yielding degenerate states $(\Delta^\dagger)^n|\psi\rangle$ for arbitrary $n$.

The above property immediately leads to the exact solution of the many-body Hamiltonian in Eq.~\eqref{eq:int:before:projection}. First, the Hamiltonian is semipositive definite. Therefore, finding a state with zero energy signifies the ground state. Second, a trivial ground state is the vacuum state $| vac\rangle$. In the absence of fermions, the density operator becomes zero, which qualifies it as a ground state of $H_{int}$. Third, by taking advantage of the aforementioned SU(2) symmetry, $(\Delta^\dagger)^n| vac\rangle$ is also a degenerate ground state of the system. In fact, for any smooth function $f(x)$, $f(\Delta^\dagger)| vac\rangle$ is a ground state of the many-body Hamiltonian. In simpler terms, any quantum state consisting of any number of Cooper pairs, but devoid of unpaired electrons, is a ground state of $H=H_{int}$.

Moreover, as shown in Ref.~\cite{Xu2022}, the system has a single-particle gap, that is, the introduction of unpaired electrons costs a finite amount of energy. This single-particle gap, as well as the single particle Green's function of Bogoliubov quasi-particles, are also exactly solvable, indicating that the magnitude of the single-particle gap is approximately the projected interaction strength $\tilde{V}^{eff}$, which corresponds to the scale of $V^{eff}_q$ multiplied by the $\lambda$ values as defined in Eq. \eqref{eq:density:projected}~\cite{Xu2022}.

Because Cooper pairs cost no energy, while unpaired single electrons are gapped, at $T=0$, all electrons will form Cooper pairs. 
Furthermore, the single-particle gap defines a temperature scale $T^*\sim \tilde{V}^{eff}$, below which thermal fluctuations are incapable of breaking Cooper pairs, that is, a pseudogap phase (Fig.~\ref{fig:phasediagram}). This has been directly verified by QMC simulations~\cite{Xu2022}.

\mysection{Is the system a superconductor?}
The answer to this question is negative in the flat-band limit. 
This is because the superconductivity order parameter is a generator of the aforementioned SU(2) symmetry in the flat-band limit. According to the Mermin-Wagner theorem, symmetry breaking within the SU(2) group cannot manifest in 2D at any finite temperature, consequently yielding a zero $T_c$. In other words, the quasi-long-range order for phase coherence, essential for 2D superfluids/superconductors, is unattainable at any finite temperature in 2D in the flat-band limit. This deduction agrees with the results of the QMC simulations.

To attain a finite $T_c$ in this model, a deviation from the exact flat-band limit is necessary. Introduction of a finite bandwidth $W\ne 0$ into the first term of Eq. \eqref{eq:full:Ham} explicitly disrupts the SU(2) symmetry. Consequently, the superconducting state only spontaneously breaks the U(1) charge symmetry, a process enabling quasi-long-range order in 2D via a Berezinskii-Kosterlitz-Thouless transition. In this regime, although an exact solution is unavailable, QMC simulations~\cite{Xu2022} affirm the emergence of a superconducting dome (Fig.~\ref{fig:phasediagram}), with the optimum $T_c$ arising near half-filling and the value of the optimum $T_c$ being comparable with the bandwidth of the flat band $T_c \sim W$, assuming that the bandwidth $W$ remains smaller than the projected interaction energy scale $\tilde{V}^{eff}$.

Inequality $W < \tilde{V}^{eff}$ also implies $T_c < T^*$. Ergo, above the superconducting dome, a non-superconducting pseudogap phase consistently emerges within the range $T_c<T<T^*$, where bound states of Cooper pairs are formed, but phase coherence fails to develop. The emergence of a pseudogap phase is beyond the weak-coupling theory or the BCS approximation. This divergence highlights the distinctive nature of this proposed experimental configuration relative to other polariton-induced superconducting states expounded in prior research.

\mysection{Discussion.}
Lastly, we estimate $T^*$ and $T_c$ of the proposed experimental configuration. 
In the weak correlation regime where $\tilde{V}^{eff}\ll W$,  $T_c  \propto \exp(-W/\tilde{V}^{eff})$ is exponentially suppressed. In contrast, our system realizes the strong correlation limit where $\tilde{V}^{eff}\gg W$. In this limit,  $T^*\sim \tilde{V}^{eff}$ and $T_c\sim W$. These relations are found to be remarkably universal, insensitive to other control parameters or microscopic details~\cite{Xu2022}. 

Due to the absence of exponential suppression, $T^*$ and $T_c$ may reach higher values than the typical $T_c$ of BCS superconductors. QMC simulations for a model system of $W\sim 0.06$~meV yield a $T_c\sim 0.13$~meV$\sim 2W$~\cite{Xu2022}. For typical twisted TMD bilayers, the flat band's bandwidth hovers around a few meV. Therefore, a cautiously optimistic estimate places $T_c$ at several tens of Kelvin. The condensate-induced attraction $\tilde{V}^{eff}$ can be controlled by engineering the polariton modes and polariton condensate order parameter, potentially reaching $10s$~ meV~\cite{Cotlet2016,Villegas2019,Julku2022}, or $T^*$ over $100$~K. 

Importantly, even without attaining a high $T_c$, the robust prediction of a substantial temperature range between $T_c$ and $T^*$ has significant implications. Within this range, Cooper pairs form, but do not condense into a superconducting state. This quantum phase is a non-Fermi liquid state, whose thermodynamic quantities deviate strongly from the Fermi liquid theory, for example, the violation of the Wiedemann-Franz law and the absence of quantum oscillations. With the predictability of the solvable model and controllability of moir\'e structures and polariton systems, the proposed setup may provide an effective platform for emulating the pseudogap phenomena seen in other strongly-correlated high-temperature superconductors.
\\

	\begin{acknowledgments}
		{\it Acknowledgments} K.S. and H.D. acknowledge support by the National Science Foundation (DMR 2132470), the Office of Naval Research (N00014-21-1-2770), and the Gordon and Betty Moore Foundation (N031710).

  \end{acknowledgments}

\bibliography{TMD.bib}

\renewcommand{\thesection}{S-\arabic{section}}
\renewcommand{\theequation}{S\arabic{equation}}
\setcounter{equation}{0}  
\renewcommand{\thefigure}{S\arabic{figure}}
\setcounter{figure}{0}  

\vspace{12pt}

\appendix
{\centering
{\large {\bf Appendix}}}
\vspace{-12pt}

\section{\label{app:H} A. Model Hamiltonian}
For the polaritons, $H_0^p$ is defined as:
\begin{align}
H_0^p=\sum_\mathbf{k} 
\begin{pmatrix}
a_c^\dagger & a_{ex}^\dagger
\end{pmatrix}
\begin{pmatrix}
\epsilon_{k}^c & g_0
\\
g_0 & \epsilon_{k}^e
\end{pmatrix}
\begin{pmatrix}
a_c \\ a_{ex}
\end{pmatrix}
\end{align}
where $a_c$ and $a_{ex}$ represent the annihilation operators for the cavity photon and exciton, respectively. Their coupling strength is denoted by $g_0$.
For electrons in twisted bilayer TMDs, the $k \cdot p$ Hamiltonian from Ref.~\cite{MacDonald2019} is adopted:
\begin{align}
H_0^e=\sum_\mathbf{k,\tau} 
\begin{pmatrix}
\psi_{t,\tau}^\dagger & \psi_{b,\tau}^\dagger
\end{pmatrix}
\mathcal{H}_{\tau,k}
\begin{pmatrix}
\psi_{t,\tau} \\ \psi_{b,\tau}
\end{pmatrix}
\end{align}
where $\psi_{t,\tau}$ and $\psi_{b,\tau}$ are fermionic operators for electrons in the top and bottom layers, respectively. Here, $\tau =\pm K$ denotes the valley index. The Hamiltonian kernel is
\begin{align}
\mathcal{H}_{\tau=+K,k}
=\begin{pmatrix}
- \frac{\hbar^2 (k- K_t)^2}{2 m^*}+\Delta_b(r) & \Delta_T(r)
\\
\Delta_T(r) & - \frac{\hbar^2 (k- K_b)^2}{2 m^*} +\Delta_t(r)
\end{pmatrix}
\end{align}
while the $\mathcal{H}_{tau=-K}$ is the time-reversal of $\mathcal{H}_{tau=+K}$.
$K_t$ ($K_b$) is the wavevector of the K point in the top (bottom) layer.
$\Delta_b$, $\Delta_t$ and $\Delta_T$ are the moir\'e potentials induced by the bottom and top layers and the tunneling strength between the two layers. The control parameters in this moir\'e Hamiltonian have been computed via first-principle techniques for various TMD materials (See examples in \cite{MacDonald2019, Devakul2021,PhysRevB.107.L201109, Wang2023}). 

Interactions between electrons and excitons, as well as between excitons, are approximated as short-ranged, resulting in
\begin{align}
H_{I}^{e-ex}=&\frac{g_{e-ex}}{2}\sum_{q}  \rho_{q} \rho_{ex,-q}
\\
H_{I}^{ex-ex}=&\frac{g_{ex-ex}}{2}\sum_{q}  \rho_{ex,q}\rho_{ex,-q}
\end{align}
where $\rho_{ex,q}$ is the exciton density at wavevector $q$.

\section{B. Partial cancellation of intravalley interactions.}
In this section, we present a proof regarding the partial cancellation of intravalley interactions as defined by the Hamiltonian
\begin{align}
H_{intra}=\frac{1}{2}\sum_{k,k',q} V_q \psi^\dagger_{k+q,\tau}\psi^\dagger_{k'-q,\tau}\psi_{k',\tau}\psi_{k,\tau}.
\label{eq:si:intra:interaction1}
\end{align}
Utilizing the anticommutation relation  $\{\psi_{k,\tau},\psi_{k',\tau}\}=0$, we can rewrite Eq.~\eqref{eq:si:intra:interaction1} as
\begin{align}
H_{intra}=-\frac{1}{2}\sum_{k,k',q} V_q \psi^\dagger_{k+q,\tau}\psi^\dagger_{k'-q,\tau}\psi_{k,\tau}\psi_{k',\tau}.
\end{align}
Performing the substitution $q\to q-k+k'$, we have
\begin{align}
H_{intra}=-\frac{1}{2}\sum_{k,k',q} V_{q-k+k'} \psi^\dagger_{q+k',\tau}\psi^\dagger_{k-q,\tau}\psi_{k,\tau}\psi_{k',\tau}.
\end{align}
Swapping $k$ and $k'$, we obtain an alternative yet equivalent expression for the intravalley interaction defined in Eq.~\eqref{eq:si:intra:interaction1}
\begin{align}
H_{intra}=-\frac{1}{2}\sum_{k,k',q} V_{q+k-k'} \psi^\dagger_{q+k,\tau}\psi^\dagger_{k'-q,\tau}\psi_{k',\tau}\psi_{k,\tau}.
\label{eq:si:intra:interaction2}
\end{align}
The sign flip between Eq.~\eqref{eq:si:intra:interaction1} and Eq.~\eqref{eq:si:intra:interaction2} suggests a hint towards cancellation. By averaging Eqs.~eqref{eq:si:intra:interaction1} and~\eqref{eq:si:intra:interaction1}, we arrive at
\begin{align}
H_{intra}=\frac{1}{2}\sum_{k,k',q} \frac{V_q-V_{q+k-k'}}{2} \psi^\dagger_{q+k,\tau}\psi^\dagger_{k'-q,\tau}\psi_{k',\tau}\psi_{k,\tau}.
\end{align}
This allows us to introduce a new intravalley interaction strength, denoted as $V_{intra}$
\begin{align}
V_{intra}= \frac{V_q-V_{q+k-k'}}{2}
\end{align}
and rewrite $H_{intra}$ as
\begin{align}
H_{intra}=\frac{1}{2} \sum_{k,k',q} V_{intra}
\psi^\dagger_{q+k,\tau}\psi^\dagger_{k'-q,\tau}\psi_{k',\tau}\psi_{k,\tau}.
\end{align}
$V_{intra}$ now depends on $q$, $k$ and $k'$, and notably, it vanishes as $k-k'$ approaches zero. For small $k-k'$, $V_{intra}$ behaves as
\begin{align}
V_{intra}= \partial_q V_q |k-k'| +O(|k-k'|^2)\sim \textrm{const} |k-k'|.
\end{align}
In the extreme case of $k-k'=0$, $V_{intra}=0$, aligning with the Pauli exclusion principle, which requires $\psi^\dagger_{k+q,\tau}\psi^\dagger_{k'-q,\tau}\psi_{k',\tau}\psi_{k,\tau}$ to vanish at $k-k'=0$.

At low doping (small $k_F$), because interactions are  predominantly governed by terms with $k-k'<2k_F$, we can take the small $k-k'$ approximation, and thus 
\begin{align}
V_{intra}\propto |k-k'| \propto \sqrt{\rho_e},
\end{align}
where $\rho_e$ denotes the fermion density.

\end{document}